\documentclass[a4paper]{jpconf}
\usepackage{amsmath,amssymb,latexsym,graphicx,multirow,psfrag,cite}

\newcommand{\Is}{\mathcal{I}}

\newcommand{\Ns}{\mathcal{N}}
\newcommand{\Rs}{\mathcal{R}}
\newcommand{\Ss}{\mathcal{S}}

\allowdisplaybreaks

\begin{document}
\title{Gravitational radiation from compact binaries in scalar-tensor gravity}
\author{R N Lang}
\address{Department of Physics, University of Florida, Gainesville, Florida 32611, USA}
\ead{lang@phys.ufl.edu}

\begin{abstract}
General relativity (GR) has been extensively tested in the solar system and in binary pulsars, but never in the strong-field, dynamical regime. Soon, gravitational-wave (GW) detectors like Advanced LIGO and eLISA will be able to probe this regime by measuring GWs from inspiraling and merging compact binaries. One particularly interesting alternative to GR is scalar-tensor gravity. We present progress in the calculation of second post-Newtonian (2PN) gravitational waveforms for inspiraling compact binaries in a general class of scalar-tensor theories. The waveforms are constructed using a standard GR method known as ``direct integration of the relaxed Einstein equations,'' appropriately adapted to the scalar-tensor case. We find that differences from general relativity can be characterized by a reasonably small number of parameters. Among the differences are new hereditary terms which depend on the past history of the source. In one special case, binary black hole systems, we find that the waveform is indistinguishable from that of general relativity.  In another, mixed black hole-neutron star systems, all differences from GR can be characterized by only a single parameter. 
\end{abstract}

\section{Introduction}
\label{sec:intro}
2015 marks the centennial of Einstein's theory of general relativity (GR).  In the last 100 years, GR has withstood every test thrown at it.  Nevertheless, we still expect GR to fail at some energy scale, since we have no consistent quantum theory of gravity.  Modifications to gravity may also explain the phenomena of dark matter and dark energy.  It is therefore important that we keep testing Einstein's theory as it moves into its second century.  Notably, general relativity has not yet been tested in the strong-field, dynamical regime that exists in coalescing compact binaries made of neutron stars and/or black holes.  These systems are the primary targets of current and future gravitational-wave (GW) detectors, such as Advanced LIGO and Advanced Virgo on the ground and eLISA in space.  By carefully measuring the gravitational waveform of a compact binary with these detectors, and comparing it to templates from different theories, we can set limits on deviations from general relativity or possibly discover new physics.

A particularly important alternative to GR is the collection of scalar-tensor theories of gravity \cite{fm03}.  These theories have a long history, dating back over 50 years, and while they are perhaps an overly simple modification to Einstein's theory (with just the addition of some scalar fields to GR's metric tensor $g_{\mu\nu}$), they remain well-motivated.  For instance, many so-called $f(R)$ theories, which may explain the acceleration of the universe without dark energy, can be expressed in the form of a scalar-tensor theory \cite{dt10}.  Scalar-tensor theories are also potential low-energy limits of string theory \cite{fm03}.

Most analyses of gravitational waveforms in alternative theories of gravity include only the lowest-order effects in the waveform model. However, real templates used by gravitational-wave detectors must be calculated to high post-Newtonian (PN) order, an expansion in powers of the source velocity $v/c$.  In this paper, we report on progress in calculating gravitational waveforms in a subclass of scalar-tensor theories to second post-Newtonian (2PN) order.  The details of this work can be found in papers by Mirshekari and Will \cite{mw13} (hereafter MW) and Lang \cite{l14} (hereafter L14).  Throughout the paper, we use units in which $c = 1$.  Greek indices run from 0-3, while Latin indices run from 1-3.  Repeated indices are summed over.

We are specifically interested in theories with a single, massless scalar field $\phi$.  In the simplest and most familiar scalar-tensor theory, known as Brans-Dicke theory, the coupling function $\omega(\phi)$ is a constant; here, however, we allow it to be any function with a Taylor expansion.  We work in the ``Jordan frame,'' in which the scalar field does not couple directly to the matter (though, of course, $\phi$ will indirectly impact the matter through its role in determining the metric).  With all of these assumptions, the scalar-tensor field equations are given by

\begin{subequations}
\begin{align}
G_{\mu \nu} &= \frac{8\pi}{\phi}T_{\mu \nu}+\frac{\omega(\phi)}{\phi^2}\left(\phi_{,\mu}\phi_{,\nu}-\frac{1}{2}g_{\mu\nu}\phi_{,\lambda}\phi^{,\lambda}\right)+\frac{1}{\phi}(\phi_{;\mu\nu}-g_{\mu\nu}\Box_g \phi) \, ,
\label{eq:tensoreqn}\\
\Box_g \phi &= \frac{1}{3+2\omega(\phi)}\left(8\pi T - 16\pi\phi\frac{\partial T}{\partial \phi}-\frac{d\omega}{d\phi}\phi_{,\lambda}\phi^{,\lambda}\right) \, ,
\label{eq:scalareqn}
\end{align}
\end{subequations}
where $G_{\mu \nu}$ is the Einstein tensor, $T_{\mu \nu}$ is the stress energy of matter and nongravitational fields, and $T \equiv g^{\alpha \beta}T_{\alpha \beta}$ is its trace.  We use commas to denote ordinary derivatives.  Semicolons denote covariant derivatives (taken using $g_{\mu \nu}$ in the usual way), and $\Box_g \equiv g^{\alpha \beta}\partial_\alpha \partial_\beta$.

Since scalar-tensor theories do not obey the strong equivalence principle, the motion and gravitational-wave emission of a binary depend on the internal composition of its constituent bodies.  To handle this complication, we have adopted the approach of Eardley \cite{e75}.  We treat the matter stress-energy tensor as a sum of delta functions located at the position of each compact object.  However, instead of assigning each body a constant mass, we let the mass be a function of the scalar field, $M_A = M_A(\phi)$.  This gives the matter an indirect dependence on $\phi$ [hence the term $\partial T/\partial \phi$ in \eqref{eq:scalareqn}], even though we still work in the Jordan frame.  In the final results, this dependence will appear as the ``sensitivity'' of the mass to variations in the scalar field,

\begin{equation}
s_A \equiv \left(\frac{d \ln M_A(\phi)}{d \ln \phi}\right)_0 \, ,
\end{equation}
as well as derivatives of this quantity.  (The subscript 0 means that the derivative should be evaluated using the asymptotic value of the scalar field, $\phi_0$.)  In the weak-field limit, the sensitivity is proportional to the Newtonian self-gravitational energy per unit mass of the body.  For neutron stars, the sensitivity depends on the mass and equation of state of the star, with typical values 0.1--0.3 \cite{wz89}.  For black holes, $s = 0.5$, and all derivatives vanish.

\section{Solving the field equations}

To solve the field equations, we use a method known as direct integration of the relaxed Einstein equations (DIRE).  While DIRE was originally developed for GR by Will, Wiseman, and Pati \cite{w92, ww96, pw00, pw02}, it can be easily adapted to scalar-tensor theory if the Einstein equations are replaced by the field equations \eqref{eq:tensoreqn}-\eqref{eq:scalareqn}.

The first step is to introduce new variables which will help us place the field equations in a ``relaxed'' form.  We assume that far away from the sources, the metric reduces to the Minkowski metric, $\eta_{\mu \nu}$, and the scalar field tends to a constant $\phi_0$.   We introduce a rescaled scalar field,

\begin{equation}
\varphi \equiv \frac{\phi}{\phi_0} \, .
\label{eq:varphi}
\end{equation}
Next, we define a conformally transformed metric, $\tilde{g}_{\mu \nu} \equiv \varphi g_{\mu \nu}$, with determinant $\tilde{g}$, and a ``gravitational field,''

\begin{equation}
\tilde{h}^{\mu \nu} \equiv \eta^{\mu \nu} - \sqrt{-\tilde{g}}\tilde{g}^{\mu \nu} \, .
\label{eq:htilde}
\end{equation}
We use a tilde to differentiate this field from the gravitational field defined in general relativity, $h^{\mu \nu}$, which has the same definition but with the normal metric $g_{\mu \nu}$ in place of the conformally transformed metric $\tilde{g}_{\mu \nu}$.  [See, for instance, (2.2) in \cite{ww96}.]  We impose the Lorenz gauge condition

\begin{equation}
{\tilde{h}^{\mu \nu}}_{\hphantom{\mu\nu}, \nu} = 0 \, .
\end{equation}
Then the field equation \eqref{eq:tensoreqn} reduces to

\begin{equation}
\Box_\eta \tilde{h}^{\mu \nu} = -16\pi \tau^{\mu \nu} \, ,
\label{eq:tensorwave}
\end{equation}
where $\Box_\eta \equiv \eta^{\alpha \beta}\partial_\alpha \partial_\beta$ is the flat-spacetime wave operator and the source is

\begin{equation}
\tau^{\mu\nu} \equiv (-g)\frac{\varphi}{\phi_0}T^{\mu \nu} + \frac{1}{16\pi}(\Lambda^{\mu\nu} + \Lambda_s^{\mu \nu}) \, .
\label{eq:taumunu}
\end{equation}
Here $g$ is the determinant of $g_{\mu \nu}$.  The first term of \eqref{eq:taumunu} represents the {\em compact} piece of the source.  The quantity $\Lambda^{\mu\nu}$ represents the gravitational-field contribution to the stress energy.  It is constructed so that it has the same functional form as in GR (except with components of $\tilde{h}^{\mu \nu}$ taking the place of components of $h^{\mu \nu}$).  The quantity $\Lambda_s^{\mu \nu}$ is an entirely new contribution to the stress energy which depends on $\varphi$.  Both $\Lambda^{\mu \nu}$ and $\Lambda_s^{\mu \nu}$ are given explicitly in L14, Eqs.\ (2.9)-(2.11).  The scalar field equation \eqref{eq:scalareqn} can also be written as a flat-spacetime wave equation,

\begin{equation}
\Box_\eta \varphi = -8\pi \tau_s \, ,
\label{eq:scalarwave}
\end{equation}
with source

\begin{equation}
\tau_s \equiv -\frac{1}{3+2\omega}\sqrt{-g}\frac{\varphi}{\phi_0}\left(T-2\varphi\frac{\partial T}{\partial \varphi}\right)-\frac{1}{8\pi}\tilde{h}^{\alpha\beta}\varphi_{,\alpha\beta}+\frac{1}{16\pi}\frac{d}{d\varphi}\left[\ln\left(\frac{3+2\omega}{\varphi^2}\right)\right]\varphi_{,\alpha}\varphi_{,\beta}\sqrt{-\tilde{g}}\tilde{g}^{\alpha\beta} \, .
\label{eq:scalarsource}
\end{equation}

The wave equations \eqref{eq:tensorwave} and \eqref{eq:scalarwave} can be solved formally in all spacetime by using a retarded Green's function,

\begin{subequations}
\begin{align}
\tilde{h}^{\mu \nu}(t,\mathbf{x}) &= 4\int \frac{\tau^{\mu \nu}(t',\mathbf{x}')\delta(t'-t+|\mathbf{x}-\mathbf{x}'|)}{|\mathbf{x}-\mathbf{x}'|}d^4x' \, ,\label{eq:hintegral}\\
\varphi(t,\mathbf{x}) &= 2\int \frac{\tau_s(t',\mathbf{x}')\delta(t'-t+|\mathbf{x}-\mathbf{x}'|)}{|\mathbf{x}-\mathbf{x}'|}d^4x' \, .
\label{eq:phiintegral}
\end{align}
\end{subequations}
The delta function in both these equations restricts the integration to being over the past flat-spacetime null cone emanating from the field point $(t,\mathbf{x})$.  To obtain explicit solutions, we divide the spacetime into two regions.  The {\em near zone} is defined as the area with $|\mathbf{x}| = R < \Rs$, where $\Rs$ is the characteristic wavelength of gravitational radiation from the system.  Everything outside the near zone ($R > \Rs$) is the {\em radiation zone}.

We carry out the integrals \eqref{eq:hintegral} and \eqref{eq:phiintegral} differently depending on the location of the field point $\mathbf{x}$ of interest (near zone or radiation zone).  Once the field point is chosen, we then evaluate each integral in two separate pieces: one integral over source points $\mathbf{x'}$ in the near zone and another over source points in the radiation zone.  This makes for a total of four different integration techniques, described in detail in \cite{ww96} and L14.  It has been shown (in GR) that all terms dependent on the boundary radius $\Rs$ in one-half of the integral are exactly canceled by pieces in the other half of the integral, leaving the final answer, as expected, independent of this arbitrary parameter \cite{ww96,pw00}.  In our work, we simply assume this property and ignore any terms which depend on $\Rs$.

\section{Equations of motion}

The first step in finding the gravitational waveform is to calculate the equations of motion for the compact binary.  To do so, we must evaluate $\tilde{h}^{\mu \nu}$ and $\varphi$ at field points in the near zone, where they influence the bodies.  As stated above, this usually involves an integration over source points in both zones.  It turns out, however, that only near-zone points contribute at the order to which we work.

The calculation of the near-zone fields is an iterative one.  We begin by finding the lowest order pieces of $\tilde{h}^{00}$ and $\varphi$, which depend only on the compact piece of the source.  These fields are then plugged back into \eqref{eq:taumunu} and \eqref{eq:scalarsource} to generate the next-highest-order source.  The process continues, alternating between calculating fields and plugging them back into the source expressions, until the fields are known to the desired post-Newtonian order.  The details and results of this procedure can be found in MW.  

Once we have the fields, we can use \eqref{eq:varphi} and \eqref{eq:htilde} to solve for the metric $g_{\mu \nu}$.  From the metric, we calculate Christoffel symbols just as in GR.  The final equations of motion are a bit trickier, however, due to the violation of the strong equivalence principle.  Using the Bianchi identity, we can write the equations of motion as

\begin{equation}
T^{\mu\nu}_{\hphantom{\mu\nu};\nu}=\frac{\partial T}{\partial \phi}\phi^{,\mu} \, .
\end{equation}
A more useful form, involving the Christoffel symbols, is given in MW, Eq.\ (2.5).  We plug in for the symbols, integrate over the extent of each body, and then convert to relative coordinates.  The final 2.5PN equations of motion can be written as

\begin{equation}
\begin{split}
a^i &= -\frac{G\alpha m}{r^2}\hat{n}^i+\frac{G\alpha m}{r^2}(A_\text{PN}\hat{n}^i+B_\text{PN}\dot{r}v^i)+\frac{8}{5}\eta\frac{(G\alpha m)^2}{r^3}(A_\text{1.5PN}\dot{r}\hat{n}^i-B_\text{1.5PN}v^i) \\
&\quad +\frac{G\alpha m}{r^2}(A_\text{2PN}\hat{n}^i+B_\text{2PN}\dot{r}v^i) \, .
\label{eq:EOM}
\end{split}
\end{equation}
Here $\mathbf{\hat{n}}$ is a unit vector pointing from body 2 to body 1, $r$ is the orbital separation, $\mathbf{v}$ is the relative velocity, $\mathbf{a}$ is the relative acceleration, $m \equiv m_1+m_2$, and $\eta \equiv m_1m_2/m^2$.  The $A$ and $B$ coefficients contain all the details; they are given in MW.  We have also defined

\begin{equation}
G \equiv \frac{1}{\phi_0}\frac{4+2\omega_0}{3+2\omega_0} \, ,
\end{equation}
where $\omega_0 \equiv \omega(\phi_0)$.  This definition is chosen so that $g_{00} = -1+2GU$ matches the expression in GR for a perfect fluid (i.e., with all sensitivities equal to zero).  However, with this definition, the coupling in the equations of motion is not simply $G$, but rather $G\alpha$, with $\alpha \equiv 1-\zeta + \zeta(1-2s_1)(1-2s_2)$ and $\zeta \equiv 1/(4+2\omega_0)$.

The even orders in the equations of motion (0PN, 1PN, and 2PN) correspond to conservative parts of the motion.  These equations can be derived from a Lagrangian and admit conserved energy and momentum quantities; for details, see MW.  The odd terms (1.5PN and 2.5PN) represent radiation-reaction terms which dissipate energy and cause the binary to coalesce.  In GR, radiation reaction begins at 2.5PN order with quadrupole radiation; the 1.5PN component in our results is due to the well-known phenomenon of dipole radiation.

\section{Calculating the gravitational waves}

To date, we have calculated the tensor gravitational waves $\tilde{h}^{ij}$ to 2PN order.  The calculation requires integrating \eqref{eq:hintegral} for field points in the radiation zone.  To simplify matters, we need only be concerned with the results in a subset of the radiation zone, the {\em far-away zone}, where $R >> \Rs$.  Unlike for the equations of motion, we must consider source points in both the near and radiation zones.  

Taking the limit of large $R$, the near-zone contribution to the tensor GWs, $\tilde{h}_\Ns^{ij}$, is given by

\begin{equation}
\tilde{h}_\Ns ^{ij}(t,\mathbf{x}) = \frac{2}{R}\frac{d^2}{dt^2}\sum_{m=0}^\infty\hat{N}^{k_1}\cdots\hat{N}^{k_m}I_{\text{EW}}^{ijk_1\cdots k_m}(\tau) \, ,
\label{eq:EWwaveform}
\end{equation}
where $\mathbf{\hat{N}} \equiv \mathbf{x}/R$ is the direction from the source to the detector and $\tau \equiv t-R$ is the retarded time.  The quantities $I_{\text{EW}}^{M+2}$ are known as ``Epstein-Wagoner'' (EW) moments \cite{ew75}.  They make use of the near-zone fields calculated in MW and involve a variety of different integrals.  It is useful to consider the post-Newtonian orders of the various EW moments.  The lowest order moment, $I_\text{EW}^{ij}$, has a quadrupole character.  As in general relativity, we define its lowest order piece to be 0PN order.  It also has higher-order contributions, specifically at 1PN, 1.5PN, and 2PN orders. When substituted into \eqref{eq:EWwaveform}, it generates GWs at all of these orders.

The three-index (octupole) moment $I_\text{EW}^{ijk}$ begins at 0.5PN order but also comprises pieces at 1.5PN and 2PN orders.  When plugged into \eqref{eq:EWwaveform}, it generates GWs at all of these orders.  Each successive EW moment begins one-half post-Newtonian order higher than the previous moment.  The final moment we need is the six-index moment $I_\text{EW}^{ijklmn}$, which consists only of a 2PN contribution (and generates 2PN gravitational waves).

Once they have been calculated to the desired order for an arbitrary number of bodies, the EW moments are simplified to the two-body case of interest and rewritten in relative coordinates.  Finally, they are substituted back into \eqref{eq:EWwaveform}, which involves taking time derivatives.  Whenever an acceleration appears in one of these derivatives, we substitute in the equations of motion \eqref{eq:EOM}. 

The tensor waves also include contributions from radiation-zone source points.  The first step to finding them is to derive an expression for the source $\tau^{ij}$ in the radiation zone.  Since there are no compact sources at $R > \Rs$, $\tau^{ij}$ is composed purely of field terms.  The details of this derivation, and of the subsequent calculation of the GWs, are found in L14.

It turns out that the radiation-zone integrals produce GWs at 1.5PN and 2PN orders.  For the sake of clarity, we separate GWs generated by $\Lambda^{ij}$ from those generated by $\Lambda_s^{ij}$.  The former produces terms which differ from the GR versions only by a scaling factor.  For instance, at 1.5PN order, we find the contribution 

\begin{equation}
\tilde{h}^{ij} = \frac{4G(1-\zeta)m}{R}\left[\frac{11}{12}\dddot{\Is}^{ij}+\int_0^\infty ds\; \overset{(4)}{\Is}\vphantom{\Is}^{ij}(\tau-s)\ln \frac{s}{2R+s}\right] \, .
\label{eq:tail1}
\end{equation}
Similar terms exist at 2PN order.  The moment $\Is^{ij}$ is defined in L14, Eq.\ (6.3a).  The first term in \eqref{eq:tail1} is {\em instantaneous}, depending only on the state of the binary at the current (retarded) time.  All near-zone contributions to the waveform are also instantaneous.  The second term in \eqref{eq:tail1}, however, is {\em hereditary}, depending on the entire past history of the source.  Specifically, the presence of the logarithm means that this is a ``tail'' term, arising physically from the scattering of the waves off the background spacetime.  This phenomenon is present in GR at this order; scalar-tensor theory only modifies the coefficient.

The field terms $\Lambda_s^{ij}$ produce new pieces of the gravitational waveform not seen in GR.  For instance, at 1.5PN order, we have
\begin{equation}
\tilde{h}^{ij} = \frac{4G(1-\zeta)m_s}{R}\left(-\frac{1}{12}\dddot{\Is}_s^{ij}\right)+ \frac{4}{R}\frac{1-\zeta}{\zeta}\left(\frac{1}{6}\int_{-\infty}^\tau \ddot{\Is}_s^i(\tau')\ddot{\Is}_s^j(\tau')d\tau'-\frac{1}{6}\dot{\Is}_s^{(i}\ddot{\Is}_s^{j)}-\frac{1}{18}\Is_s^{(i}\dddot{\Is}_s^{j)}\right) \, ,
\label{eq:memory1}
\end{equation}
where $m_s$, $\Is_s^i$, and $\Is_s^{ij}$ are defined in L14, Eqs.\ (6.10), (6.11a), and (6.11b).  Similar terms exist at 2PN order.  Eq.\ \eqref{eq:memory1} contains a hereditary contribution which is {\em not} a tail.  Terms like these---hereditary but without logarithms---do appear in GR, but not until 2.5PN order. Their appearance beginning at 1.5PN order in scalar-tensor theory results from the nonvanishing of the scalar dipole moment, $\Is_s^{i}$, the same effect which causes radiation reaction to appear at 1.5PN order in the equations of motion.  The 1.5PN hereditary integral can be described as a ``gravitational-wave memory'' term.  It contains a zero-frequency, or DC, component, in addition to the usual oscillatory terms.  (The new 2PN hereditary integral, not shown, does not.)  

\section{Discussion}
To find the final tensor waveform, we add the contributions from the near and radiation zones.  It can be written as a post-Newtonian expansion,

\begin{equation}
\tilde{h}^{ij} = \frac{2G(1-\zeta)\mu}{R}[\tilde{Q}^{ij}+P^{1/2}Q^{ij}+PQ^{ij}+P^{3/2}Q^{ij}+P^2Q^{ij}+O(\epsilon^{5/2})]_\text{TT} \, ,
\end{equation}
where the superscripts on $P$ denote the PN order of each term. ``TT'' denotes the transverse-traceless projection; to save time, we throw away pieces at any stage of the calculation which will not survive this final projection.  Each term in the expansion can be found in L14; note that there the 1.5PN and 2PN terms have been divided into the contributions from the near zone and the contributions from the radiation zone [e.g., \eqref{eq:tail1} and \eqref{eq:memory1}].

The expressions are significantly more complicated than the GR equivalents, Eq.\ (6.11) of \cite{ww96}.  Nevertheless, all deviations can be characterized by a small number of parameters, including $G$, $\zeta$, $\alpha$, and several others ($\bar{\gamma}$, $\bar{\beta}_i$, $\bar{\delta}_i$, $\bar{\chi}_i$, $\Ss_+$, and $\Ss_-$), given in or near MW Table I.  All of these parameters, as expected, depend on the fundamental parameters of the theory (Taylor coefficients of the coupling function $\omega(\phi)$ and the asymptotic scalar field $\phi_0$) and/or the sensitivities and sensitivity derivatives of the sources.  The same set of parameters characterizes the deviations of the equations of motion from the GR version.  

In general, the tensor waves in scalar-tensor theory have the same general functional form as the GR waves, with only the addition of these parameters to the coefficients.  In some cases, however, the existence of a nonvanishing scalar dipole moment generates entirely new terms.  One example is the 1.5PN contribution from the quadrupole moment; in GR, this moment only contributes at 0PN, 1PN, and 2PN orders.  Others include the new hereditary integrals.  

For binary black holes, the expressions simplify considerably.  The 2.5PN equations of motion and 2PN tensor waves turn out to have the exact same forms as in general relativity if the masses are rescaled by a factor $1/(1-\zeta)$.  Since the masses of the bodies are defined by their Keplerian motion, this rescaling is unmeasurable.  Therefore, to these orders, the motion and tensor GW emission of two black holes in scalar-tensor theory are indistinguishable from the motion and emission in GR.  This result is not surprising.  Hawking showed that stationary, asymptotically flat black holes in vacuum are identical in both theories \cite{h72}, leading to conjectures that the same might be true for black hole binaries.  The result is supported by other analytic studies (to all PN orders, but only leading order in the mass ratio) \cite{ypc12} and numerical studies \cite{h11}.  Of course, it depends strongly on the assumptions we made (no potential for the scalar field, a time-independent $\phi_0$).  One unfortunate consequence of this result is that eLISA will be unlikely to provide information about these particular scalar-tensor theories; ground-based detectors, which will detect many inspirals involving neutron stars, are more suited to the task.

For a system containing one neutron star (say, body 1) and one black hole, the expressions also simplify somewhat.  In this case, the motion and waves are identical to those in GR through 1PN order (after a mass rescaling).  At 1.5PN order, deviations start to occur.  However, the deviations are always parametrized by the single parameter $Q = \zeta(1-\zeta)^{-1}(1-2s_1)^2$.  Because $Q$ contains no information on the derivatives of the coupling function $\omega(\phi)$, we cannot, at this order, formally distinguish the waveform produced in the Brans-Dicke theory from that produced in a general scalar-tensor theory of the type we consider.  One could imagine doing so by measuring masses and sensitivities of multiple neutron stars.

Work is underway on the calculation of the scalar gravitational waves.  Some of the low-order pieces can be found in \cite{w14}, Eqs.\ (101) and (102).  The process is similar to that for the tensor waves; however, there are some unique complications.  For instance, the near-zone calculation also relies on the computation of moments (like the EW moments).  In the scalar case, however, these include monopole and dipole moments.  Since we defined the quadrupole moment (and the waves it generated) as 0PN order, the monopole and dipole moments turn out to begin at -1PN and -0.5PN orders, respectively.  These moments must then be calculated to relative 3PN and 2.5PN orders in order to find the complete 2PN gravitational waves.  The number of integrals required to find the 3PN-accurate monopole moment is several times that needed for any of the tensor EW moments; furthermore, several of the integrals are extremely lengthy to compute.  Things become even more difficult when considering the energy loss from the system, ultimately needed to determine the gravitational-wave phasing.  While the 2PN energy loss from tensor waves can be found by knowing up to the 2PN form of the waves, the 2PN energy loss from scalar waves requires knowledge of certain 2.5PN pieces of the waveform.  It also requires the 3PN equations of motion, one-half PN order beyond that calculated in MW.  There is a still a great deal of work to do in calculating even the simplest scalar-tensor gravitational waveforms.

\ack

We thank the organizers of the 10th International LISA Symposium for an excellent conference.  We are grateful to Clifford Will and Saeed Mirshekari for their hard work on the first phase of this calculation and their useful insights on the rest.  This work was supported in part by the National Science Foundation, Grants No.\ PHY-0965133, No.\ PHY-1260995, and No.\ PHY-1306069.

\section*{References}
\bibliographystyle{iopart-num}
\bibliography{LISAX}

\end{document}